\documentstyle[aps,12pt,epsfig]{revtex}

\newcommand{ \be }{\begin{equation}}
\newcommand{ \ee }{\end{equation}}
\newcommand{ \bea }{\begin{eqnarray}}
\newcommand{ \eea }{\end{eqnarray}}

\begin{document}
\normalsize
\title{$\Lambda$ Production and Flow in Au+Au Collisions at 11.5A GeV/c} 

\author{
   J.~Barrette$^5$, R.~Bellwied$^{9}$,
  S.~Bennett$^{9}$, R.~Bersch$^7$, P.~Braun-Munzinger$^2$, 
  W.~C.~Chang$^7$, W.~E.~Cleland$^6$, M.~Clemen$^6$, 
  J.~Cole$^4$, T.~M.~Cormier$^{9}$, 
  Y.~Dai$^5$, G.~David$^1$, J.~Dee$^7$, O.~Dietzsch$^8$, M.~Drigert$^4$,
  K.~Filimonov$^3$, S.~C.~Johnson$^7$, 
  J.~R.~Hall$^9$, T.~K.~Hemmick$^7$, N.~Herrmann$^2$, B.~Hong$^2$, 
  Y.~Kwon$^7$,
  R.~Lacasse$^5$, Q.~Li$^{9}$, T.~W.~Ludlam$^1$,
  S.~K.~Mark$^5$, R.~Matheus$^{9}$, S.~McCorkle$^1$, J.~T.~Murgatroyd$^9$,
  D.~Mi\'{s}kowiec$^2$,
  E.~O'Brien$^1$,  
  S.~Panitkin$^7$, V.~Pantuev$^7$, P.~Paul$^7$, 
  T.~Piazza$^7$, M.~Pollack$^7$, 
  C.~Pruneau$^9$, Y. ~J. ~Qi$^5$,
  M.~N.~Rao$^7$, E.~Reber$^4$, M.~Rosati$^5$, 
  N.~C.~daSilva$^8$, S.~Sedykh$^7$, U.~Sonnadara$^6$, J.~Stachel$^3$, 
  E.~M.~Takagui$^8$, 
  V. ~Topor ~Pop$^5$, M. Trzaska$^7$,
  S.~Voloshin$^6$, T.~B.~Vongpaseuth$^7$,
  G.~Wang$^5$, J.~P.~Wessels$^3$, C.~L.~Woody$^1$, N.~Xu$^7$,
  Y.~Zhang$^7$, C.~Zou$^7$\\ 
(E877 Collaboration)
}
\address{
 $^1$ Brookhaven National Laboratory, Upton, NY 11973\\
 $^2$ Gesellschaft f\"ur Schwerionenforschung, 64291 Darmstadt, Germany\\
 $^3$ Universit\"at Heidelberg, 69120 Heidelberg, Germany\\
 $^4$ Idaho National Engineering Laboratory, Idaho Falls, ID 83402\\
 $^5$ McGill University, Montreal, Canada\\
 $^6$ University of Pittsburgh, Pittsburgh, PA 15260\\
 $^7$ SUNY, Stony Brook, NY 11794\\
 $^8$ University of S\~ao Paulo, Brazil\\
 $^9$ Wayne State University, Detroit, MI 48202\\
}

\date{July 4, 2000}
\maketitle

\vskip 0.3cm

\begin{abstract}
New data on $\Lambda$ production in Au+Au collisions at 11.5 A GeV/c 
are presented.
The measurements cover the rapidity range from $y$=2.0
to 3.5 and transverse momenta from $p{_t}$=0.15 GeV/c to
1.5 GeV/c. The rapidity distributions, 
transverse momentum spectra, and azimuthal distributions
are presented for different centralities of the collision. 
A strong positive directed flow at forward rapidity is 
observed for semicentral collisions. 
The measured spectra, yields and directed flow
are compared with the predictions of RQMD v2.3 model.    
\end{abstract}


\section{Introduction}

Strangeness production in heavy-ion collisions when compared to 
proton-proton collisions is potentially a sensitive probe of 
collective energy deposition and therefore of 
heavy ion reaction mechanisms in general. 
Its study may provide insight into 
the properties of hot and dense nuclear matter \cite{rafelski82}.
Indeed, enhanced strangeness production has been observed at AGS 
and SPS energies \cite{ahle00,antinori99}.
 
 Experimental measurements indicate that high baryon density is reached
in central heavy-ion collisions at both AGS \cite{bar95,ahle98_1} and 
SPS \cite{appel99_1}. 
Via comparisons with models that reproduce the experimental data, 
it has been concluded that one reaches 
baryon densities of up to
10 times normal nuclear matter density and energy densities of the order of 
2 GeV/fm$^3$ in the center of the fireball \cite{johana98}. 
These values are well into the range where, based 
on lattice QCD calculations, one expects a 
baryon-rich deconfined phase. The realization 
\cite{munzinger96,heinz99_1d} that fireballs with saturated 
strangeness cannot be produced in purely hadronic scenario
has lent strong support to the interpretation that a deconfined (at least 
partly) phase has been found in central 
ultrarelativistic nuclear collisions.
Consequently it is of great importance to investigate 
further aspects of strangeness production, such as, e. g., 
flow effects.  

The experimental and theoretical studies of collective flow of 
various types have been an important component of the investigation of
relativistic heavy-ion collisions  
\cite{ritter98,miklos99,fwang99_1,cassing99_1}.
The experimental data on transverse collective flow
phenomena at Bevalac and SIS  energies (1-2A GeV) \cite{ritter98}, 
AGS (4-11A GeV) \cite{bar94,bar97_1,bar97_2,pink99}
and SPS energies (150-200A GeV) 
\cite{wienold96,bear97,appel98_2,ceretto98},
have led to a renewed intense theoretical interest in this topic
\cite{ollit98,risc96,sorge97,sorge99,heinz99_1,levy99}.
It was pointed out
that collective flow directly probes the equation of state of 
nuclear matter and can provide a useful information about the 
possible phase transition \cite{risc96,sorge99}.

In order to obtain the information about 
the  nuclear equation of state, 
systematic studies of  
 freeze-out observables as a function of collision system size and 
bombarding energy are necessary \cite{miklos99,fwang99_1,cassing99_1}.
These observables include strangeness and antibaryon production.
Since the bulk of strangeness produced is
carried by kaons, measurements of kaon spectra and their flow 
in heavy-ion collisions have been
systematically carried out at different energies 
\cite{ritter98,senger99,ogilvie99}
and their in-medium properties have been investigated 
\cite{brown98}.

Experimental studies of $\Lambda$ and 
 $\bar{\Lambda}$ production and flow are now available at
SIS/GSI \cite{ritman95,justice98} 
and at AGS energies 
\cite{eiseman94,ahmad96,wu96,yang96,best97_2,ahmad98,chang99,alex99}. 
At SIS energies it was found that the $\Lambda$ mean field 
constructed on the basis of the quark model leads to 
a good description of the observed  
in-plane transverse flow of $\Lambda$'s \cite{faessler99}.
Differences between the $K^+$ and $\Lambda$ 
flow due to their different mean field potential in dense matter
have also been discussed \cite{brown99_2}: while the $\Lambda$ flow is
basically similar to that of nucleons, the $K^+$ flow almost
 disappears. Recently, directed flow of neutral strange particles 
 in heavy-ion collisions at AGS has been investigated in a relativistic
 transport model (ART)  \cite{bin99_2},
showing that the smaller $\Lambda$ directed flow relative to the proton
 flow can be accounted for by a weaker mean-field potential as in  
the constituent quark model and from hypernuclei phenomenology.
Therefore, a detailed study of the directed flow of strange particles in 
conjunction with other variables should help in understanding the 
relative importance of different reaction mechanisms. 

 In this paper we present new 
experimental results on $\Lambda$ production 
obtained from the 1995 run of the E877 experiment. 
In section 2, we briefly describe the experimental setup. 
In section 3, the analysis steps for $\Lambda$ identification 
 are discussed. The experimental results, 
including the rapidity distributions, invariant mass spectra, 
and azimuthal distributions, and their comparison with RQMD v2.3 model 
are presented in section 4. 
Discussion and conclusion will be given in the last section.


\section{E877 apparatus}

\begin{figure} [hbt!]
\centering
\epsfig{file=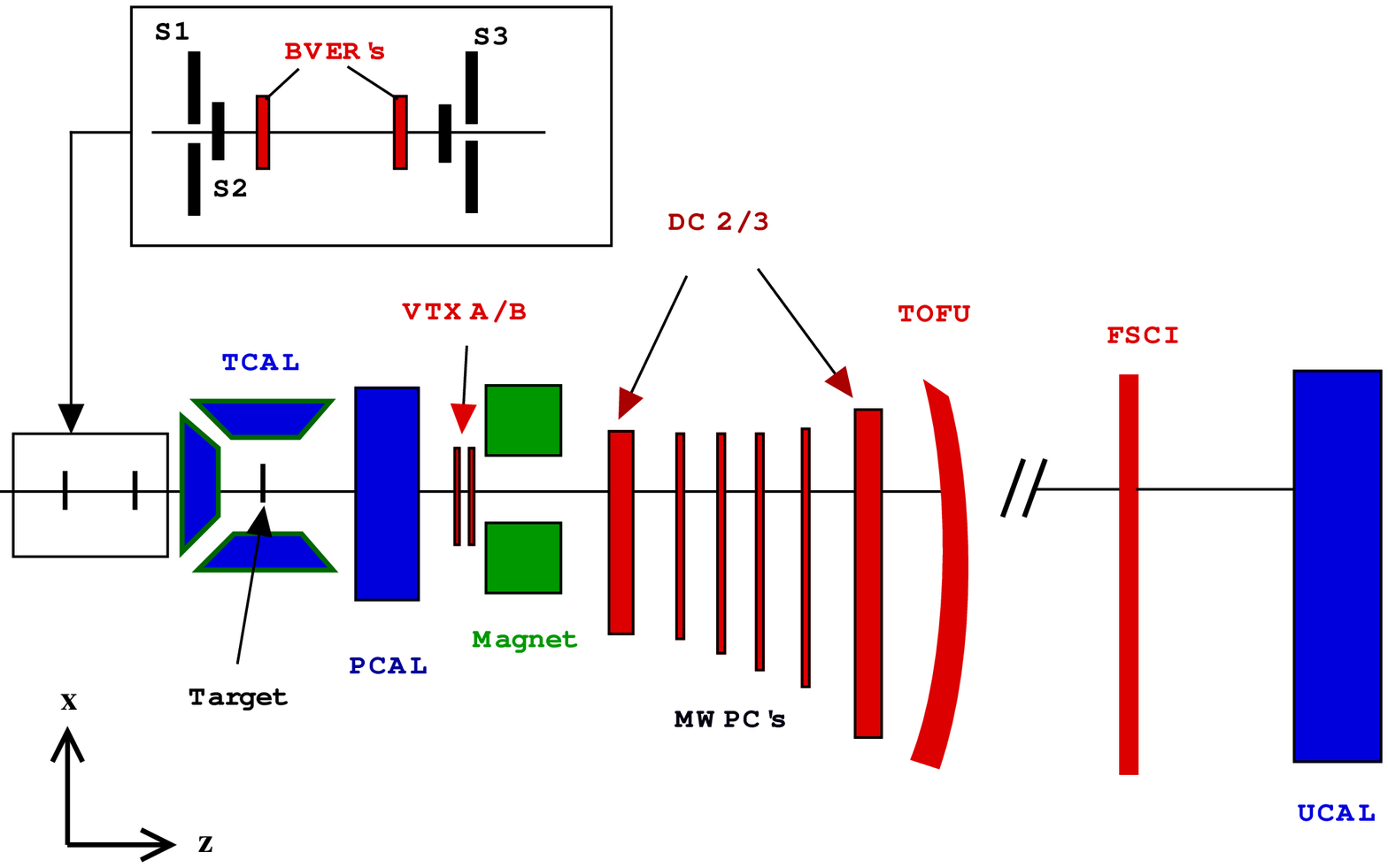,height=10cm, width=14cm} 
 \caption[E877 experimental setup]{\small
Schematic view of the E877 experimental setup for the 1995 run. Au
beam is incident from the left.
\label{fig:setup}}
\end{figure}

 A schematic view of the E877 apparatus is shown in
Fig.~\ref{fig:setup}.
For the 1995 run, the silicon beam vertex detectors (BVER's) 
were upgraded from single-sided
silicon wafers with one-dimensional pitch of 50 $\mu$m to  
double-sided wafers with a 200 $\mu$m pitch in both the 
$x$ and $y$ directions (see Fig.~\ref{fig:setup}). Using these detectors 
the  position and angle of beam particles at the target were determined 
with an accuracy of 300 $\mu$m in coordinates and 60 $\mu$rad in angle. 
The centrality of the collision and reaction
plane orientation were obtained from the transverse energy 
distribution measured in
the target calorimeter (TCAL), and participant calorimeter (PCAL).  Both
calorimeters had $2 \pi$ azimuthal coverage and combined,
provided nearly complete polar angle ($\theta$) coverage: 
TCAL and PCAL covered the
pseudorapidity regions $-0.5 <\eta <0.8$ and $0.8 <\eta < 4.2$, 
respectively, where $\eta = -\rm{ln[\tan(\theta/2)]}$.
The centrality selection was quantified by the ratio
$\sigma_{top}$/$\sigma_{geom}$, 
where $\sigma_{top}={\int_{E_t}^{\infty}} (d\sigma/dE_t')dE_t'$ and 
$\sigma_{geom}$ is the geometrical cross section of 
the colliding Au+Au nuclei.
The analysis of anisotropic transverse collective flow requires 
the precise determination of the reaction plane which is defined by 
the impact parameter vector and the beam axis.
The reaction plane orientation was determined event by event from 
the azimuthal anisotropy in the transverse energy distribution \cite{ydai99}. 
The reaction plane resolution was evaluated by studying the correlation
between flow angles measured in different pseudorapidity intervals 
\cite{vol96,ydai99}.  

Charged particles emitted in the forward direction and passing through a
collimator ($ -134\,\,$mrad  $< \theta_{horizontal} < 16 $ mrad, 
$ -11\,\,$ mrad$ < \theta_{vertical} < 11\,\, $ mrad) 
were analyzed in a high resolution magnetic spectrometer.  
The spectrometer acceptance covered mostly the
forward rapidity region.  The momentum of each particle was measured
using two drift chambers, DC2 and DC3, 
whose pattern recognition capability was aided by four
multi-wire proportional chambers (MWPC).  The average momentum
resolution was $\Delta p/p \approx$ 3\% limited by multiple scattering.  A
time-of-flight hodoscope (TOFU) located directly behind the tracking
chambers provided the time-of-flight with an average resolution of
85~ps \cite{lac98}.  Energy loss information from TOFU 
was used to determine the particle charge.

\begin{figure} [hbt!]
\centering
\epsfig{file=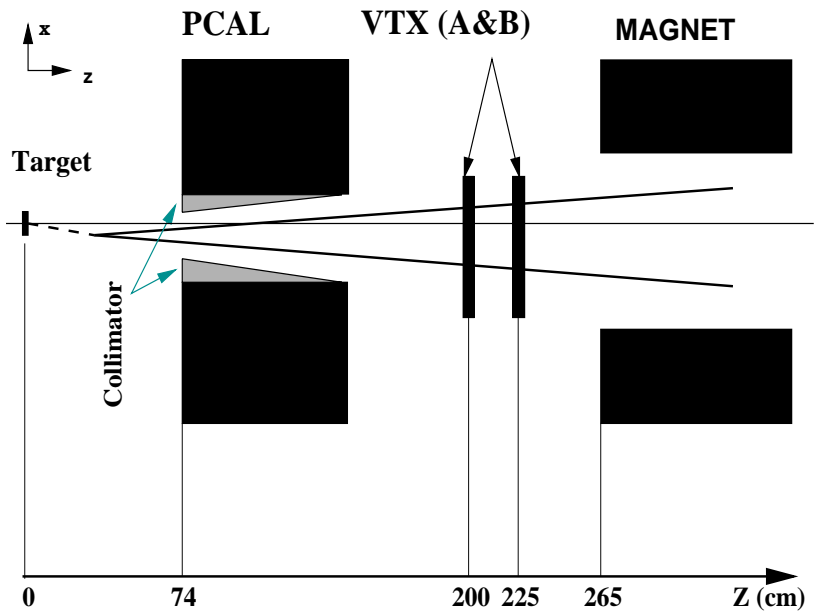,height=8cm,width=11cm,clip=} 
 \caption[Schematic layout of upstream detectors of the spectrometer]{\small
Schematic layout of upstream detectors of the spectrometer.
\label{fig:vtx_setup}}
\end{figure}

For the 1995 run, two identical
highly segmented cathode pad detectors (VTX) were instrumented and installed
between the PCAL/collimator and the spectrometer magnet
(Fig.~\ref{fig:vtx_setup}).  They provided
a precise measurement of the x-coordinate (bending plane coordinates) 
of the track before deflection in the magnetic field. 
This allowed to reconstruct the decay vertices of the 
particles such as $\Lambda$ baryons and improve the
signal-to-background ratio for identification of rare
particles such as $K^-$ and antiprotons.
The active area of each pad chamber consisted of 10 rows of chevron
shape pads with each row having 53 pads and one anode
wire placed above each row. The position resolution along the
x-direction was about 300 $\mu$m, while the resolution in y-direction
was determined by the
wire spacing (5 mm). 
A detailed description of the design,
implementation and performance of the vertex chambers can be found in
\cite{bersch95}.


\section{Data analysis}

In our data analysis $\Lambda$'s were identified by reconstructing  
($p$,$\pi^- $) pairs from
their characteristic $V_0$ decay topology 
$ \Lambda \rightarrow p + \pi^-$.
Single tracks were first reconstructed by the standard E877 tracking program 
``Quanah''~\cite{bar94_1} which assumed that particle originates  
from the target. 
The invariant mass of ($p$,~$\pi^- $) pair was  calculated using
 $\Lambda$-decay
kinematic hypothesis only and was used for the $\Lambda$ identification. 
To separate the $\Lambda$'s from a large combinatorial background of the
directly produced $p$ and $\pi^-$, we performed the following analysis
steps. 

From the information provided by the VTX detectors,
which are not part of the standard E877 tracking program, 
we determined the track segments upstream of the spectrometer magnet 
for $p$ and $\pi^-$ and
reconstructed the assumed decay vertex
($V_x, V_y, V_z$) of the pair by finding
the crossing point of the $p$ and $\pi^-$ tracks. 
Since the tracking detectors had a better position
resolution in the bend plane (x-z plane) of the spectrometer, 
only the values $V_x$ and $V_z$ were used for the elimination of the
combinatorial background.

A set of conditions was applied to reduce the background. First,
it was required that the z-position of the  
decay vertex was between 25 and 180 cm 
downstream of the target.  
The minimum value was limited by the VTX 
chambers resolution and the maximum value was limited by 
the geometric position of the first VTX chamber. This condition 
drastically reduced the large combinatorial
background but also rejected about $28\%$ of the true $\Lambda$'s in 
our acceptance.  
The proton and pion tracks were also required to point away from the
interaction point in the target 
in the x-direction. We rejected the proton tracks which, after
projecting back to the target, were less than 4 mm away 
from the x-coordinate of the interaction point.
For pions we required the x-coordinate to be more than 
14 mm away from the interaction.
These cuts were very effective in rejecting pairs 
containing primary protons and pions 
but they also eliminated $\Lambda$'s whose decay products 
were emitted along the $\Lambda$ momentum vector.

Besides the above selections, we also required that the upstream track
segment determined by the VTX detectors and downstream track segment identified
by the tracking program matched at the center of the spectrometer magnet. A 
$3\sigma$ cut was applied on the difference in x-positions obtained from
fitting the upstream and downstream track segments.
The reconstructed momentum vector of the pair 
was also required to point back to the interaction point
within 2.0 mm ($3\sigma$ cut) in x-direction. 

\begin{figure}[hbt!]
\centering 
\epsfig{file=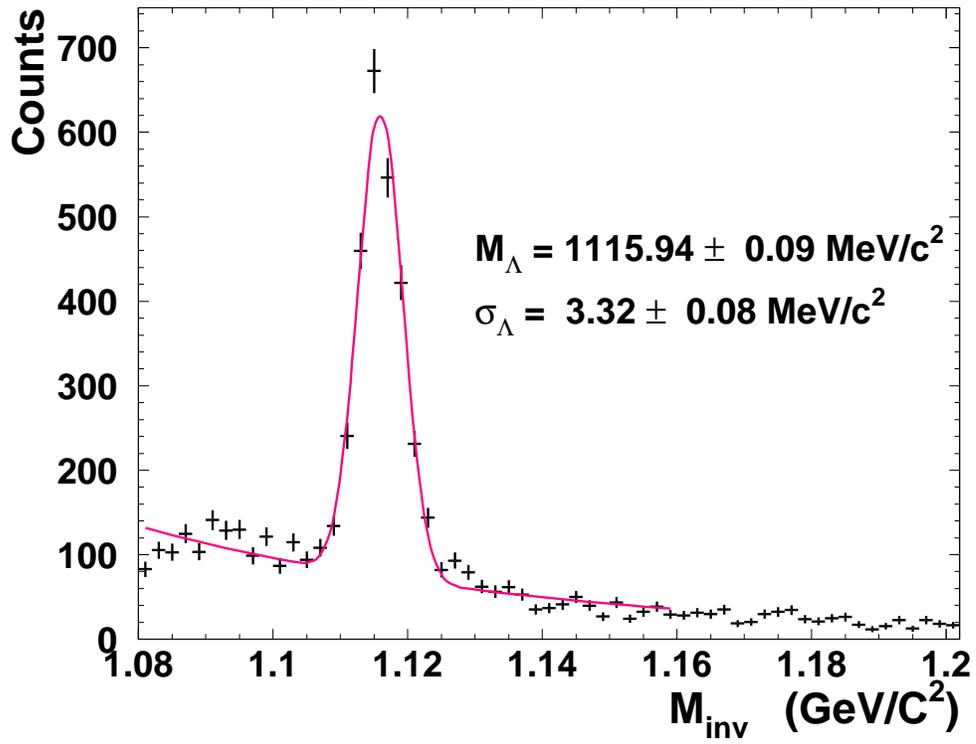,width=14cm}
 \caption[Invariant mass distribution of $p,\,\,\pi^-$ pairs]{\small
Invariant mass distribution of ($p,\pi^-$) pairs. 
\label{fig:lambda_mass}}
\end{figure}

All described cuts were optimized using a Monte Carlo
simulation \cite{yujin99}. 
After filtering by this set of conditions,  
the combinatorial background is
dramatically reduced, so the $\Lambda$ peak is well
identified in the invariant mass distribution.
A further improvement was obtained by recalculating the proton and pion
momenta using as origin the decay vertex position obtained 
from VTX detectors. This significantly improved the mass resolution of the
$\Lambda$ peak as well as the signal-to-background ratio
(see Fig.~\ref{fig:lambda_mass}). 
The remaining background under
the $\Lambda$ peak is mainly due to the
accidental combination of proton and pion tracks passing our cuts.

In order to reconstruct the $\Lambda$ spectra, the
data need to be corrected for the spectrometer acceptance
and the effects of the various conditions introduced in the analysis. 
The single track efficiency of the spectrometer downstream of the magnet 
was discussed in ~\cite{bar99_1}. 
We studied the efficiency of the upstream VTX detectors by
looking into the proton yield ratio with and without the VTX cut. 
The obtained efficiency of VTX detectors was 
about 85\% for single tracks in the
sensitive area of the detectors.

The acceptance corrections for $\Lambda$ distributions were calculated 
using a detailed Monte-Carlo simulation.
Au+Au events were generated using the RQMD v2.3 event generator
\cite{sorge95,sorge96}. The acceptance corrections on the final 
data sample were calculated 
as a function of rapidity $y$ and transverse momentum $p_t$ by
propagating the generated  particles through the E877 apparatus. All the known
effects of the spectrometer geometry, detector resolutions,
kinematics and cuts were included into the calculation. 
The acceptance for $\Lambda$-hyperons 
is of the order $10^{-3}$ for beam rapidity region
$3.0<y<3.2$, where there is relatively high reconstruction efficiency, 
and  $10^{-4}$ or less for rapidity range $2.2<y<2.5$.
The uncertainties in the estimation of the acceptance corrections,
efficiency of the VTX detectors, and background subtraction,
result in an overall systematic uncertainty of the order of 15\%
in the estimated $\Lambda$ yield.
After the background subtraction, we 
identified 2644 $\Lambda$'s from 32 millions events with centrality 
$\sigma_{top}/\sigma_{geom} < 10\,\%$. 

\begin{figure}[hbt!]
\centering 
\epsfig{file=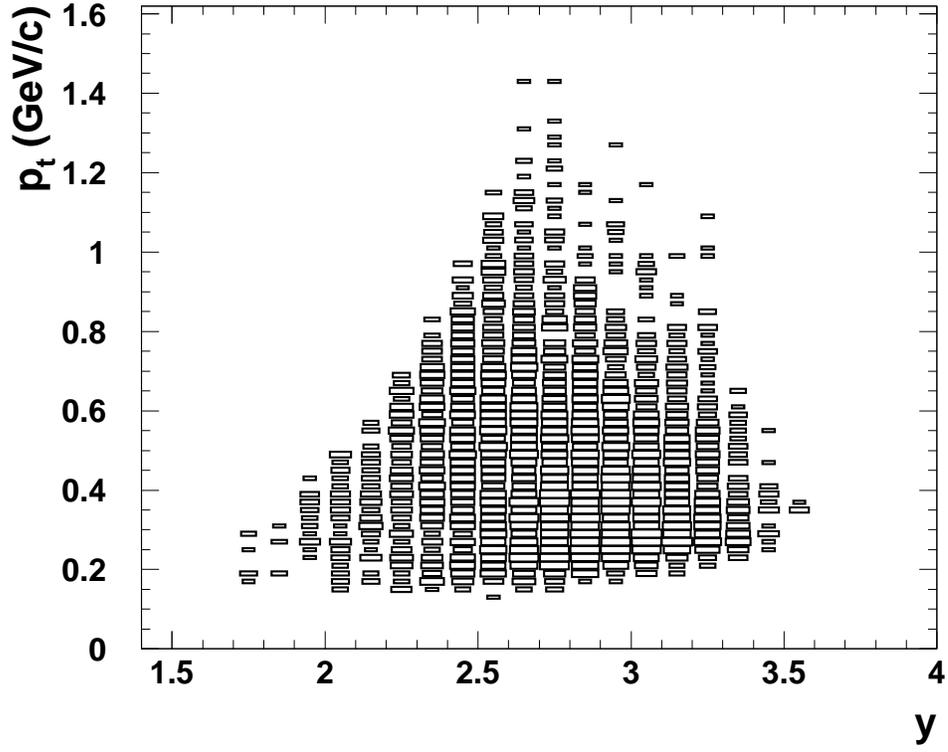,width=14cm}
 \caption[Acceptance plot for $\Lambda$'s after the pair cuts]{\small  
The measured range for $\Lambda$'s in phase space after the applied cuts.
\label{fig:sim_lam_acc}}
\end{figure}

Fig.~\ref{fig:sim_lam_acc} shows the $\Lambda$ acceptance in the ($p_t,~y$) 
phase space.
The E877 spectrometer covers a
rapidity range of $2.2<y<3.4$ and transverse momenta
$p_t$ $>$ 0.15 GeV/c. $\Lambda$'s at low $p_t$ were not reconstructed
due to the dead zones in the VTX detectors near the beam axis.  

\section{Results}

\subsection{Lambda Yield and Spectra}
 \label{sect:lam_spec}

\begin{figure}[hbt!]
\centering 
\epsfig{file=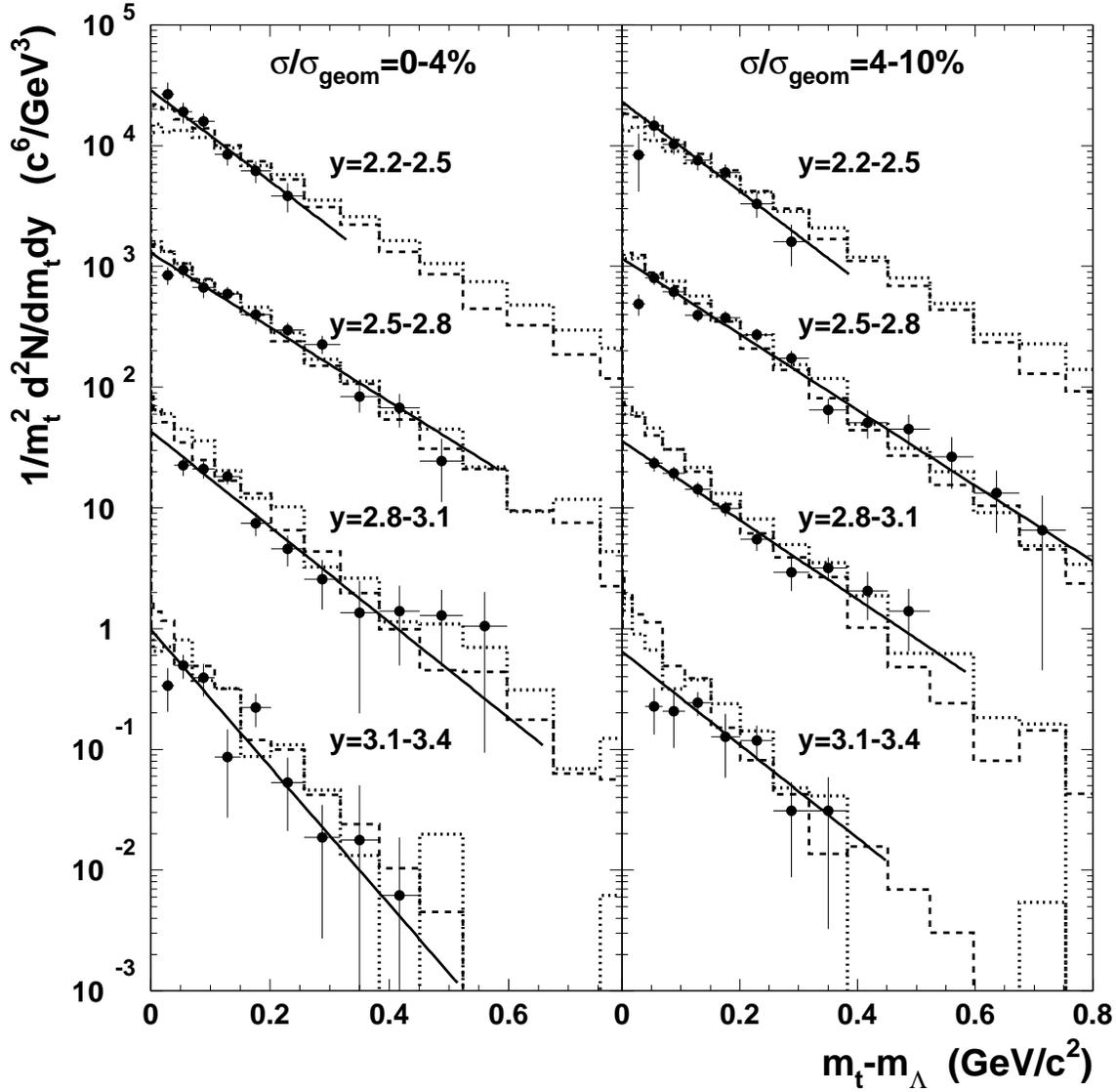,width=\textwidth}
 \caption[$\Lambda$ transverse mass spectra for the most central Au+Au
collisions]{\small   
Measured $\Lambda$ transverse mass spectra for the most central ($0-4\%\,\,
\sigma_{geom}$-left side) and semi-central
( $4-10\%\,\,\sigma_{geom}$ - right side) Au+Au collisions.
Beginning with rapidity bin $y$=3.1-3.4 spectra have been multiplied by
successive factor of 10.
The solid lines are the exponential
Boltzmann fits to the data. The dashed lines and dotted lines 
are the predictions of RQMD v2.3 model run in {\em cascade} and 
{\em mean-field} modes, respectively. 
\label{fig:l_dndmtdy}}
\end{figure}

The data were divided into
constant $p_t$ bins of 100 MeV and rapidity bins of 0.3 unit width
from $y=2.2$ to $y=3.4$. 
The lambda yield was obtained from the
invariant mass distributions in each ($y, p_t$) bin 
after corrections for acceptance, efficiency and 
background subtraction as described in section III.
In Fig.~\ref{fig:l_dndmtdy},
$\Lambda$ transverse mass spectra are presented 
for the most central collisions
($\sigma_{\rm top}/\sigma_{\rm geom}<4$\%)
and for semicentral collisions 
(4 \% $< \sigma_{\rm top}/\sigma_{\rm geom}< $ 10 \%).
The error bars include statistical errors and  
the errors from the background subtraction procedure. 
The range in $m_t$ reflects the acceptance
of the spectrometer in the different rapidity bins.
The solid lines represent the best fits to the spectra using 
Boltzmann distribution :
\begin{equation}
\frac{1}{\rm m_t^2}\cdot \frac{d^2N}{d{\rm m_t}d{\rm y}}=
\frac{N_B}{\rm m_\Lambda}\cdot\exp(-\frac{{\rm m_t}-{\rm m_\Lambda}}{T_B(y)})
\end{equation}
where $T_B(y)$ is the inverse
slope of the spectrum. The experimental $m_t$ spectra 
are in good agreement with this exponential shape.

The Relativistic Quantum Molecular Dynamics (RQMD) model \cite{sorge95}
has been widely used in describing relativistic heavy-ion collisions. 
It combines the 
classical propagation of all hadrons with string and resonance
excitations in the primary collisions of nucleons from the
target and projectile. Overlapping color strings may fuse into 
so-called ropes. Subsequently, the fragmentation products from rope,
string and resonance decays interact with each other and 
with the original nucleons, mostly via binary collisions. 
These interactions drive the system towards equilibrium \cite{sorge96}
and are responsible for the development of collective flow, even in 
the pre-equilibrium stage. 
If baryons are surrounded by other baryons they acquire effective masses.
The effective masses are generated by introducing Lorentz-invariant 
quasi-potentials into the mass-shell constraints for the momenta,
which simulates the effect of {\em mean field} \cite{sorge97}.
There are no potential-type interactions in the so-called {\em cascade mode} of
RQMD.

A comparison with RQMD v2.3 model  
run in {\em cascade} (dashed histograms) 
and {\em mean field} modes (doted histograms) is also presented in
Fig.~\ref{fig:l_dndmtdy}.
We can see that the
conventional transport hadronic model reproduces the data 
relatively well in
magnitude and shape over the measured rapidity and centrality
intervals. We note, however, especially 
in the forward rapidity region, significant (up to a factor of 2) 
differences between the cascade and mean-field modes of RQMD. 

\begin{figure}[hbt!]
\centering 
\epsfig{file=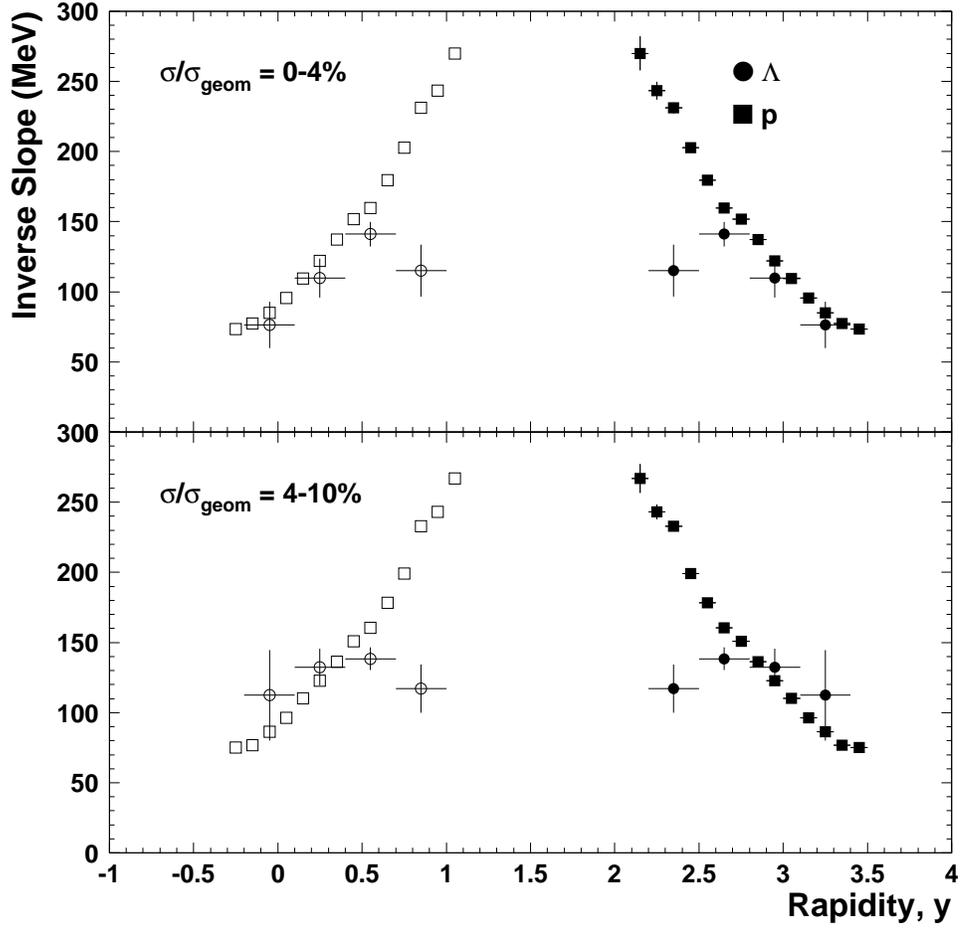,width=14cm}
 \caption[$\Lambda$ inverse slope parameters for $<4\%
\sigma_{geom}$]{\small   
The inverse slope parameters as a function of rapidity 
for central ($0-4\%\,\,\sigma_{geom}$-upper pannel) and 
semi-central ( $4-10\%\,\,\sigma_{geom}$ - lower pannel) Au+Au collisions.
The data are represented by
solid symbols and reflected about mid-rapidity (open symbols). The
proton data are extracted from the 1995 data samples.  
\label{fig:lam_slope}}
\end{figure}

The derived lambda inverse slopes $T_B(y)$
are shown in Fig.~\ref{fig:lam_slope}.
The data for $\Lambda$-hyperons  
are also compared to the slopes obtained from the
proton spectra analysis \cite{bar99_1}.
Error bars on the fit parameters are statistical only.
Except for the first data point for which the
systematic error is large due to the limited range
of the measured $m_t$ spectra,
the inverse slope parameters extracted from the
$\Lambda$ spectra are, for both centrality bins, similar to those of
the protons.
This is consistent with protons and  $\Lambda$'s
having a similar collective 
flow which is superimposed on the thermal motion at freeze-out.

\begin{figure}[hbt!]
\centering 
\epsfig{file=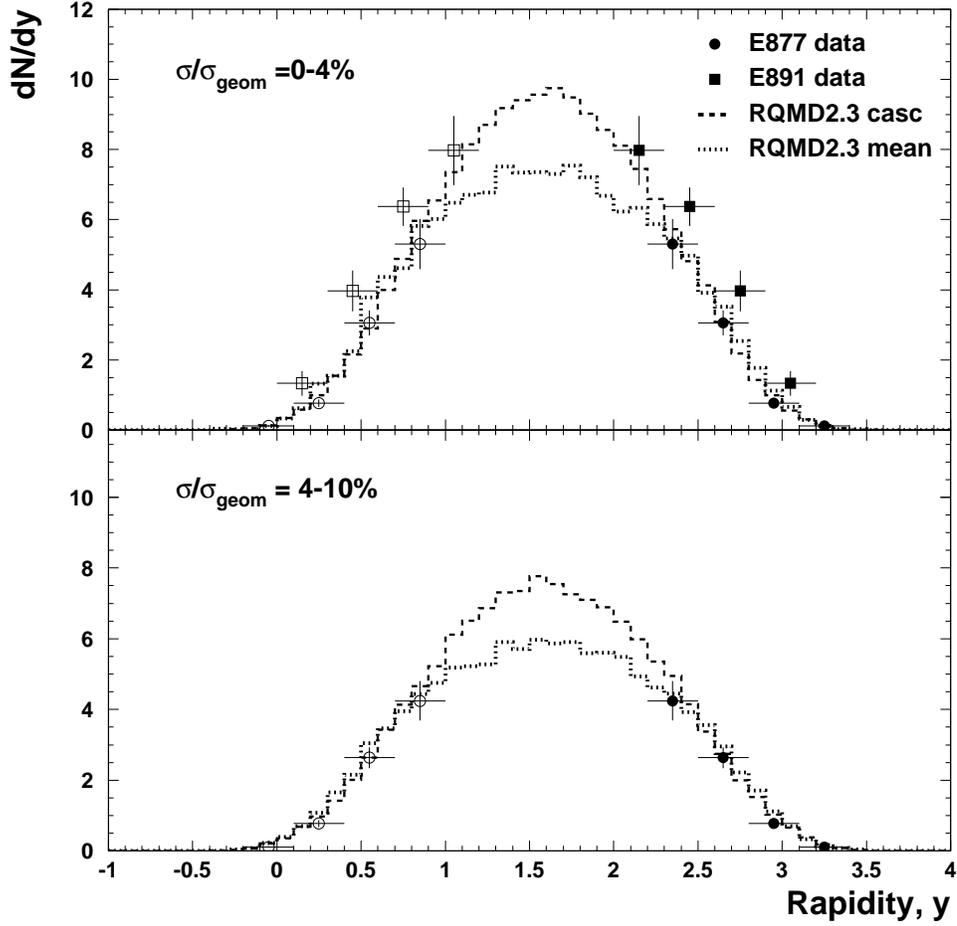,width=14cm}
 \caption[$\Lambda$ rapidity distribution for the most central Au+Au
collisions]{\small   
Lambda rapidity distribution in the most central Au+Au collisions 
($0-4\%\,\,\sigma_{geom}$- upper pannel) and 
semi-central ( $4-10\%\,\,\sigma_{geom}$ - lower pannel) Au+Au collisions.
The data are represented by solid symbols and reflected
about mid-rapidity (open symbols). The histograms correspond to 
RQMD v2.3 predictions. 
\label{fig:lam_dndy}}
\end{figure}

By integrating the transverse mass spectra where data are available  and
using the results from the Boltzmann fits to extrapolate to infinity
and to $p_t=0$ 
we obtain the $\Lambda$ rapidity distributions 
shown in Fig.~\ref{fig:lam_dndy} for the two  centrality intervals.
The data are compared with the RQMD v2.3 predictions.
The data are, within the systematic errors, in
good agreement with the model predictions. 
In the experimentally covered rapidity 
range ($2.2<y<3.4$), cascade and mean field calculations 
show a similar $\Lambda$-yield. A discrepancy between the two 
predictions of RQMD v2.3 appears in the midrapidity region, which is 
not covered by our experiment. A similar effect in RQMD calculations 
of proton yield has been recently noted ~\cite{back99}.

Fig.~\ref{fig:lam_dndy} also includes a comparison 
with the results from the E891 experiment~\cite{ahmad98} at AGS,
which covered a similar rapidity range. 
The E891 collaboration reported a yield which is roughly 
$20\,\,\%$ higher than ours. 
Such a difference is at the limit of the systematic errors of both
experiments.

\subsection{Lambda Directed Flow}
 \label{sect:lam_flow}

At lower beam energies ($<$2 A$\cdot$GeV), theoretical
studies show that directed flow of lambda
hyperons is very sensitive to the $\Lambda$ potential in dense
nuclear matter formed in heavy-ion collisions~\cite{li98}. The 
calculations also show that, at this energy, the primordial lambda hyperons  
have a weak flow as compared to nucleons. However, final-state
interactions, and especially the propagation in mean-field potentials,
enhance the lambda flow in the direction of nucleons and bring the 
theoretical results in good agreement with the experimental data from
the FOPI ~\cite{ritman95} and EOS ~\cite{justice98} collaborations.

For an emission dominated by directed flow, the azimuthal 
distribution can be parameterized by 
\begin{equation}
 \frac{1}{N_0}\,\frac{dN}{d\varphi} = 1 + 2 v_1 \cos\varphi 
\label{eqn:v1_cosfunc}
\end{equation}
where $\varphi=\phi-\psi_r$, $\phi$ is the azimuthal angle 
of the particle in the lab
frame, and $\psi_r$ is the reaction plane angle. 
The parameter $v_1$ quantifies directed flow of particles
parallel ($v_1>0$) or antiparallel ($v_1<0$) to the impact parameter
vector. 

\begin{figure}[hbt!]
\centering 
\epsfig{file=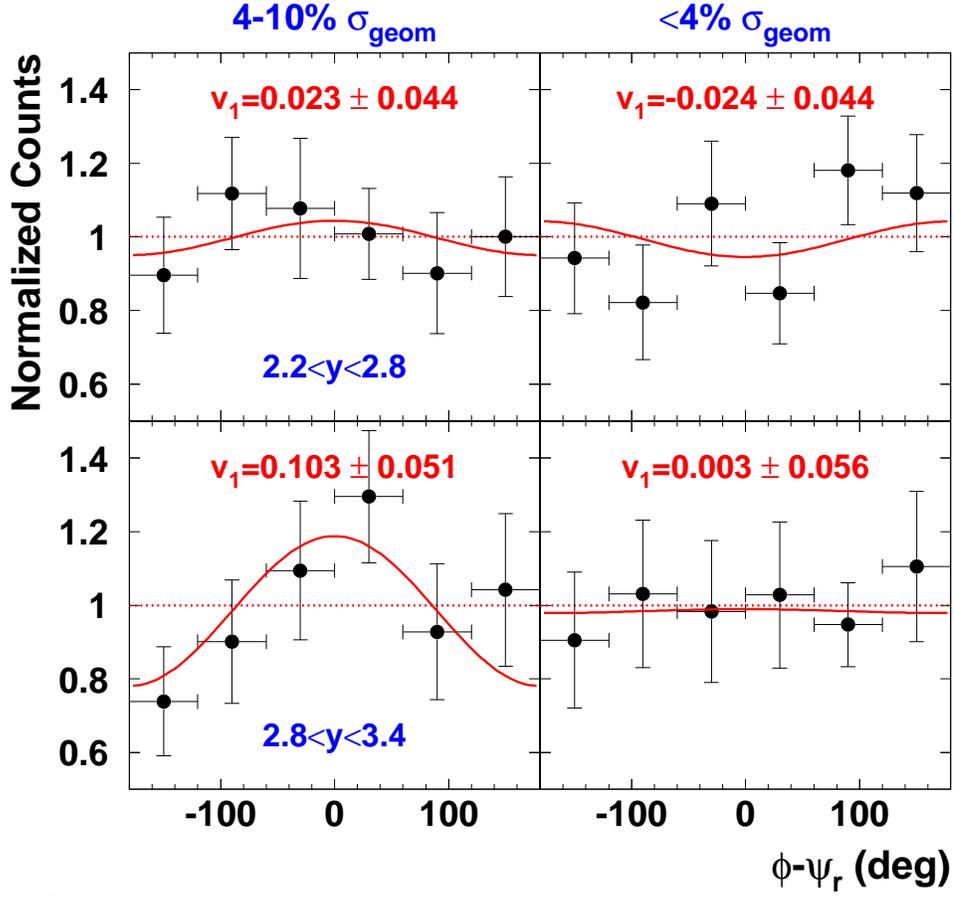,width=14cm}
 \caption[Lambda azimuthal angular distributions]{\small  
Lambda azimuthal angular distributions measured in transverse
momentum range 0.15 $<p_t<$ 1.5 GeV/c for different rapidity and
centrality bins. The distributions are normalized to unity. The solid
lines are the fits using Eq.~\ref{eqn:v1_cosfunc}. The obtained values
of $v_1$ are uncorrected for the reaction plane resolution.
\label{fig:lam_ang_v1}}
\end{figure}

We studied the azimuthal angular distribution with
respect to the reaction plane to extract the average directed flow
parameter $v_1$ estimated in the experimental $p_t$-acceptance for  
different rapidity and centrality windows. 
We divided the azimuthal angle range from $-180^{\circ}$ to $180^{\circ}$
into six equal bins.
In order to obtain the azimuthal
angular distribution of $\Lambda$'s, the $\Lambda$ yield was extracted from
the invariant mass distribution in each azimuthal angle bin after
background subtraction. The obtained 
semi-inclusive azimuthal angular distributions of
$\Lambda$'s with respect to the reaction plane are presented in
Fig.~\ref{fig:lam_ang_v1}. 
In the most central bin ($<4\% \sigma_{geom}$),
as can be expected, there is no/or very little anisotropy in the
azimuthal distributions. But in the semi-central bin ($4-10\,\,\%
\,\sigma_{geom}$), the azimuthal distributions 
exhibit a directed flow signal which becomes significant
for the forward rapidity window.

\begin{figure}[hbt!]
\centering 
\epsfig{file=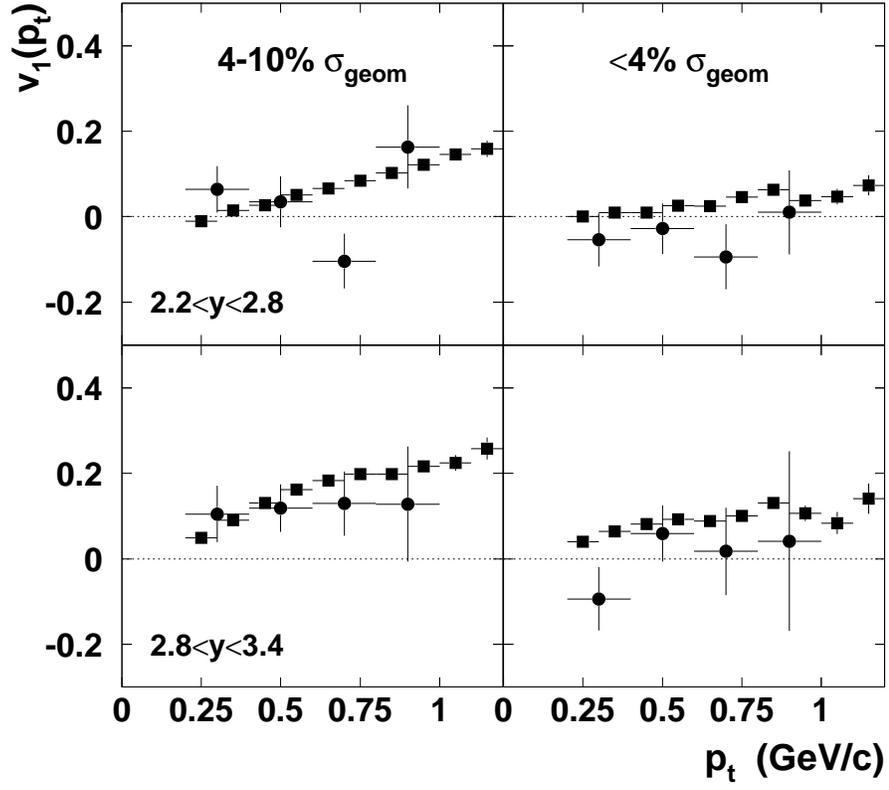,width=13cm}
\caption[Comparison of $v_1(p_t)$ data between lambdas and
protons]{\small    
Comparison of $v_1(p_t)$ data between $\Lambda$'s (solid circles) and
protons (solid squares). 
\label{fig:lam_v1pt_pro}}
\end{figure}

We also studied the differential flow ($p_t$ dependence of $v_1$) 
as a function of rapidity and centrality using the
Fourier expansion method 
which was previously used for the analysis of proton
and pion directed flow~\cite{wchang97}. 
The results are presented in Fig.~\ref{fig:lam_v1pt_pro}.
In agreement with the inclusive azimuthal angular distributions
within statistical errors  
data show no directed flow for central collisions ($<4\,\% \sigma_{geom}$).
A positive and statistically significant signal is observed  
for the semi-central collisions and forward rapidities ($2.8<y<3.4$).

\begin{figure}[hbt!]
\centering 
\epsfig{file=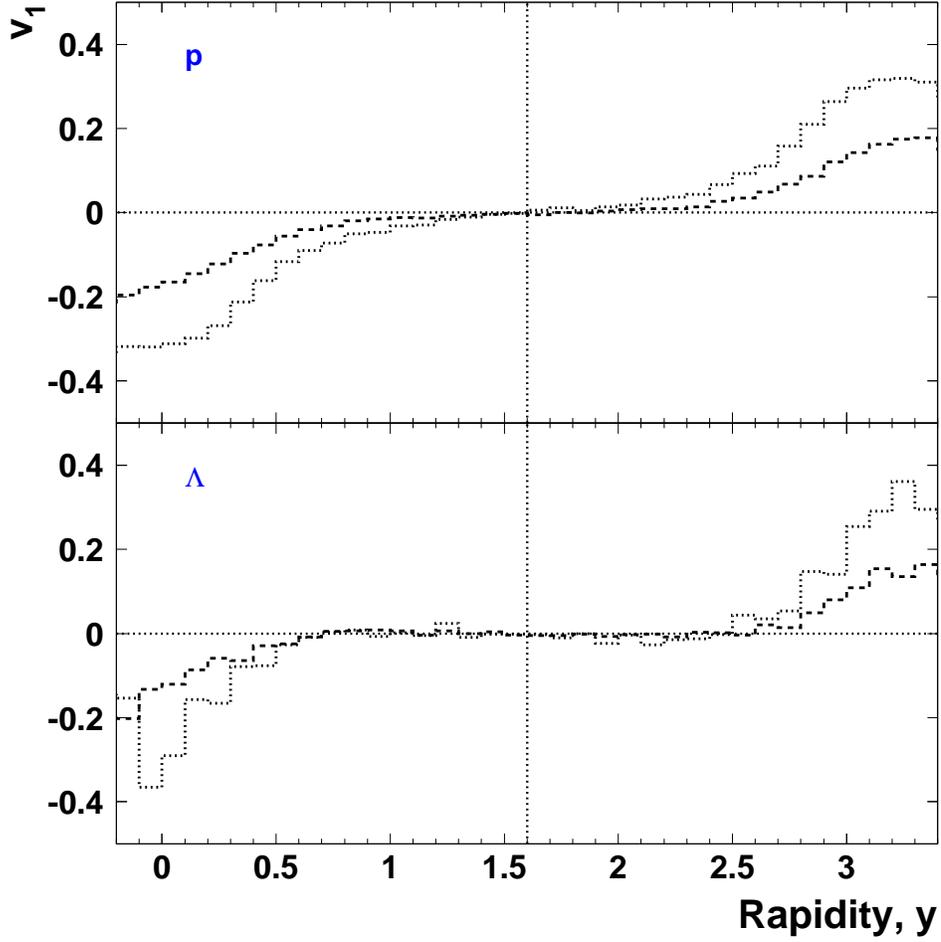,width=14cm}
\caption[Directed flows of proton and lambda as a function of
 rapidity predicted by the RQMD model (v2.3)]{\small  
Proton and lambda directed flows in Au+Au collisions ($b<$10 fm) at
11.5 A$\cdot$GeV/c as a function of rapidity predicted by the RQMD model
 (v2.3) run in {\em cascade} mode (dashed histograms) and 
{\em mean-field} mode (dotted histograms). 
\label{fig:lam_flow_rqmd}}
\end{figure}

Since nucleons are the major carriers of flow signal and directed
flow of protons has been well studied
~\cite{bar97_1,bar97_2}, it provides a good reference 
for comparison with flow of other particles.
The predictions of RQMD model v2.3 for proton and $\Lambda$ 
directed flow in Au+Au collisions are shown in
Fig.~\ref{fig:lam_flow_rqmd}.
As it has recently been shown,
RQMD well describes the amplitude of the experimentally measured 
proton flow if the effects of mean-field are included.
However, RQMD predictions  
for differential flow differ significantly from
the data, both for cascade and mean field modes  
\cite{bar97_1,bar97_2}.

In general, RQMD predicts that $\Lambda$ flow 
should be very similar to proton flow.
Both cascade and mean-field calculations  predict that 
flow of $\Lambda$'s is very small at mid-rapidity 
and that it becomes 
larger and comparable with the proton flow at $y>2.5$. This is the
region well covered by our experimental measurements. A close
inspection of the mid-rapidity region in Fig.~\ref{fig:lam_flow_rqmd}
reveals that a very small anti-flow is predicted by the
mean-field calculations with a transition from negative 
to positive flow occurring around $y=2.4$.      

\begin{figure}[hbt!]
\centering 
\epsfig{file=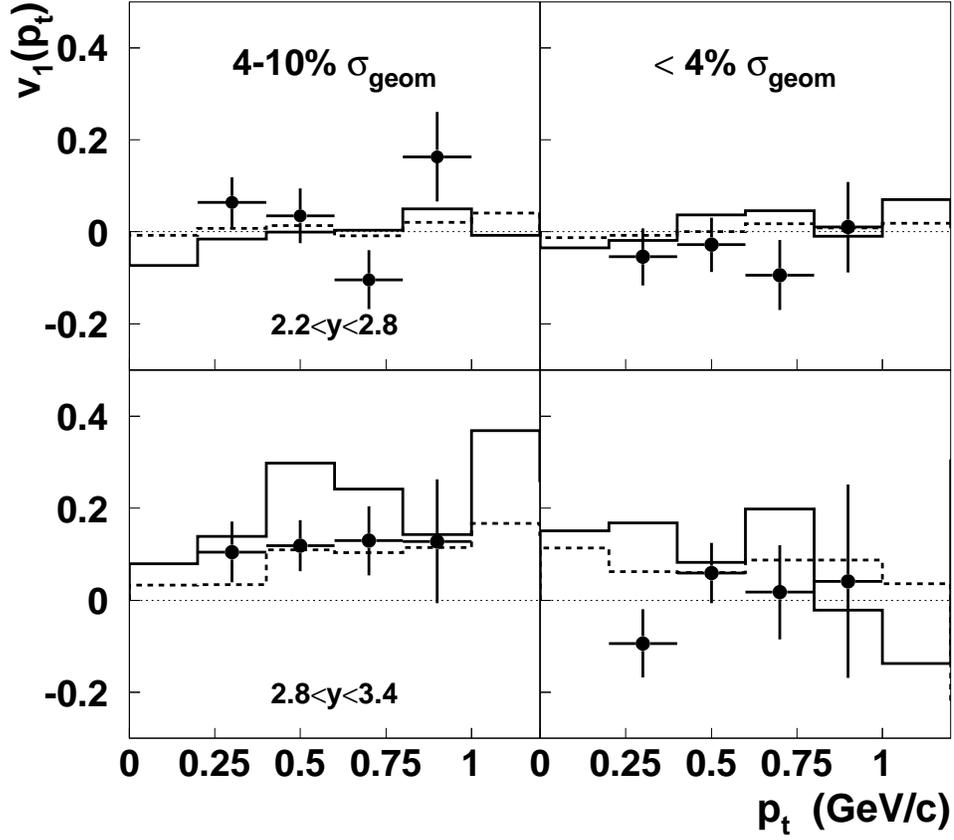,width=14cm}
 \caption[Comparison of $\Lambda$ flow data with the predictions
of RQMD v2.3 model]{\small  
Comparison of $\Lambda$ flow data (solid circles) with the predictions
of the RQMD v2.3 model, run in {\em cascade} (dashed histograms) and
{\em mean-field} (full histograms) modes. 
\label{fig:lam_v1pt_rqmd}}
\end{figure}

The comparison between the measured data and 
RQMD v2.3 predictions are presented in 
Fig.~\ref{fig:lam_v1pt_rqmd}. Although the interpretation of the
$v_1(p_t)$-dependence is limited due to low statistics, one can
conclude that the measured $p_t$ dependence of lambda directed flow is
consistent with that predicted by the model. 
The cascade calculations give a somewhat  
better description of the trend exhibited by the data.

At lower beam energies (around 2 A$\cdot$GeV), theoretical calculations
from the relativistic transport models 
indicate that mean-field potentials play a major role in 
development of $K^+$ and $\Lambda$ flows~\cite{zswang99}. 
Without any final-state
interaction, both $K^+$ and $\Lambda$ flow in the same direction as
nucleons, but with much smaller flow amplitudes. The inclusion of
rescattering with the dense matter just enhances the flow of
$K^+$ and $\Lambda$ in the direction of nucleons, as a
result of thermalization effects. However, the propagation of $K^+$
and $\Lambda$ in their mean-field potentials leads to significantly
different flow patterns for $K^+$ and $\Lambda$. Kaons are pushed away
from nucleons by their repulsive potential while lambda hyperons are
pulled towards nucleons by their attractive potential. This leads to
a predicted small anti-flow of kaons with respect to nucleons, 
and to a flow of lambda hyperons of amplitude  
very close to the flow of nucleons.

\begin{figure}[hbt!]
\centering 
\epsfig{file=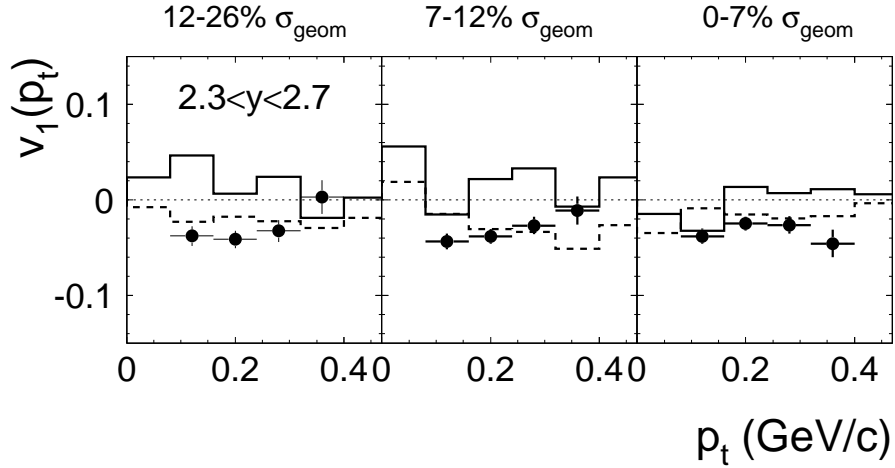,width=14cm}
 \caption[Comparison of $K^+$ flow data with the predictions of the
RQMD v2.3 model]{\small  
Comparison of $K^+$ flow data with the predictions of the RQMD v2.3
model. The histograms are the predictions of the RQMD model run in
{\em cascade} (dashed histograms) and {\em mean-field} 
(full histograms) modes. The figure is taken from~\cite{bar99_2}. 
\label{fig:k+_v1pt}}
\end{figure}

Fig.~\ref{fig:k+_v1pt} shows the comparison
of $K^+$ directed flow data~\cite{kirill99,bar99_2}
with the  RQMD v2.3 predictions.
Similar to results presented here, 
the cascade calculations give a better 
description of the data. Since $\Lambda$ production
is mainly associated with kaons in hadronic scenario, whether 
$K^+$ and $\Lambda$ medium effects persist at the AGS or
even higher energies is still an open question.


\section{Summary and Conclusion }

The lambda spectra have been measured as a function of transverse mass
and rapidity for different collision centralities. The spectra are
well described by an exponential with inverse slopes decreasing 
with increasing rapidity. 
The derived lambda inverse slopes are similar to those
obtained from the proton $m_t$ spectra. This is consistent with the 
picture of $\Lambda$'s originating from a fireball 
in local thermal equilibrium.

We have observed, for the first time at AGS, a significant 
positive directed flow of lambda hyperons at forward
rapidities (2.8$<y<$3.4) in semi-central Au+Au collisions .
The average flow amplitude for lambda hyperons is
comparable with that for protons for the same acceptance region. This
result is consistent with the measurements performed at lower 
energies ($<2$A$\cdot$GeV), and suggests that  
lambda flow follows the flow of nucleons.

The measurements 
have been compared to the predictions of RQMD model (v2.3) 
run in  {\em cascade} and in {\em mean-field} modes. 
In the covered rapidity range, the model describes 
the spectra  rather well. 
The measured differential flow ($v_1(p_t)$) is comparable to the model
predictions, with a better description 
provided by the cascade calculations.


\section*{ACKNOWLEDGMENTS}

We thank the AGS staff, W. McGahern and Dr. H. Brown for excellent
support and acknowledge the help of R. Hutter in all
technical matters. Financial support from the US DoE, the NSF, the
Canadian NSERC, and CNPq Brazil is gratefully acknowledged.

\newpage 

\newpage

\newpage

\newpage

\newpage

\newpage

\newpage

\newpage

\newpage

\newpage
\newpage

\end{document}